\documentclass[sn-mathphys,Numbered]{sn-jnl}%

\usepackage{graphicx}%
\usepackage{multirow}%
\usepackage{amsmath,amssymb,amsfonts}%
\usepackage{amsthm}%
\usepackage{mathrsfs}%
\usepackage[title]{appendix}%
\usepackage{xcolor}%
\usepackage{textcomp}%
\usepackage{manyfoot}%
\usepackage{booktabs}%
\usepackage{algorithm}%
\usepackage{algorithmicx}%
\usepackage{algpseudocode}%
\usepackage{listings}%
\usepackage{longtable}%
\usepackage{varwidth}%
\usepackage{array}%
\usepackage{pdflscape}%
\usepackage{afterpage}
\newcolumntype{P}[1]{>{\raggedright\arraybackslash}p{#1}}

\begin{document}

\title[Systematizing Modeler Experience (MX) in Model-Driven Engineering Success Stories]{Systematizing Modeler Experience (MX) in Model-Driven Engineering Success Stories}

\author[1]{\fnm{Reyhaneh} \sur{Kalantari}}\email{reyhaneh.kalantari@uottawa.ca}
\author[2]{\fnm{Julian} \sur{Oertel}}\email{julian.oertel@uni-rostock.de}
\author[3]{\fnm{Joeri} \sur{Exelmans}}\email{joeri.exelmans@uantwerpen.be}
\author[4]{\fnm{Satrio Adi} \sur{Rukmono}}\email{s.a.rukmono@tue.nl}
\author[5]{\fnm{Vasco} \sur{Amaral}}\email{vasco.amaral@fct.unl.pt}
\author[6]{\fnm{Matthias} \sur{Tichy}}\email{matthias.tichy@uni-ulm.de}
\author[7]{\fnm{Katharina} \sur{Juhnke}}\email{katharina.juhnke@zeiss.com}
\author[8]{\fnm{Jan-Philipp} \sur{Steghöfer}}\email{jan-philipp.steghoefer@xitaso.com}
\author[9]{\fnm{Silvia} \sur{Abrahão}}\email{sabrahao@dsic.upv.es}

\affil[1]{\orgdiv{Department of Engineering Design and Teaching Innovation}, \orgname{University of Ottawa}, \orgaddress{\street{800 King Edward Avenue}, \city{Ottawa}, \postcode{K1N6N5}, \country{Canada}}}
\affil[2]{\orgdiv{Department of Software Engineering}, \orgname{University of Rostock}, \orgaddress{\street{Albert-Einstein-Straße}, \city{Rostock}, \postcode{18057}, \country{Germany}}}
\affil[3]{\orgdiv{Department of Computer Science}, \orgname{University of Antwerp}, \orgaddress{\street{Middelheimlaan 1}, \postcode{2020}, \city{Antwerp}, \country{Belgium}}}
\affil[4]{\orgdiv{Department of Mathematics and Computer Science}, \orgname{Eindhoven University of Technology}, \orgaddress{\street{De Zaale}, \postcode{5600 MB}, \city{Eindhoven}, \country{The Netherlands}}}
\affil[5]{\orgdiv{Department of Computer Science}, \orgname{NOVA School of Science \& Technology}, \street{Campus de Caparica}, \postcode{2829-516} \city{Caparica}, \country{Portugal}}
\affil[6]{\orgdiv{Institute of Software Engineering and Languages}, \orgname{Ulm University}, \orgaddress{\street{James-Franck-Ring}, \postcode{89081}, \city{Ulm}, \country{Germany}}}
\affil[7]{\orgdiv{SMT-EMI6}, \orgname{Carl Zeiss SMT GmbH}, \orgaddress{\street{Rudolf-Eber-Straße 2}, \postcode{73447}, \city{Oberkochen}, \country{Germany}}}
\affil[8]{\orgname{XITASO GmbH}, \orgaddress{\street{Austraße 35}, \postcode{86153}, \city{Augsburg}, \country{Germany}}}
\affil[9]{\orgdiv{Instituto Universitario Mixto de Tecnología Informática}, \orgname{Universitat Politècnica de València}, \street{Camino de Vera s/n}, \postcode{46022} \city{Valencia}, \country{Spain}}

\abstract{Modeling is often associated with complex and heavy tooling, leading to a negative perception among practitioners. However, alternative paradigms, such as everything-as-code or low-code, are gaining acceptance due to their perceived ease of use. This paper explores the dichotomy between these perceptions through the lens of ``modeler experience'' (MX). MX includes factors such as user experience, motivation, integration, collaboration \& versioning and language complexity. We examine the relationships between these factors and their impact on different modeling usage scenarios. Our findings highlight the importance of considering MX when understanding how developers interact with modeling tools and the complexities of modeling and associated tooling.}


\maketitle

\section{Introduction}
\label{sec:introduction}

Model-driven engineering (MDE) is recognized as an established approach for developing complex software systems \cite{mussbacher2014relevance} bringing advantages during software system's development.  However, its adoption is currently hindered by a range of factors~\cite{Whittle2013}, including poor tool support and its usability \cite{Pourali2018}, social and organizational issues, and a mismatch between technical and research requirements \cite{Whittle2013}. 

As developers engage in the intricate task of modeling, whether for data, systems, or simulations, evidence from practice shows that their experiences diverge from conventional academic practices or guidelines. In \cite{abrahao2017user}, the authors first introduce the term User eXperience (UX) for MDE or \emph{Modeler Experience (MX)}. MX goes beyond the traditional definition of usability to include experience that includes individual feelings, such as emotions, affects, motivations, and values in the process of modeling, playing a crucial role in the success and adoption of MDE approaches. This paper takes the initial steps towards developing a theory of MX, as called for in the original work by \citet{abrahao2017user}.

Taking the above definition one step further, we argue that MX is not only about the modeling language, tools or individual perspectives, but also about how modeling is embedded in the organization and the mindset of the individuals. It can, therefore, be a tool to address the mindset barriers reported by Kalantari and Lethbridge~\cite{kalantari2023unveiling}. 

A fundamental insight is that modeler experience depends on the concrete context and circumstances in which modeling is used. This is in line with the principles of Context-Driven Software Engineering, where \citet{BriandBNPS17} advocate for the importance of considering contextual factors, whether human (e.g., modelers background and experience), organizational (e.g., time and cost constraints) or domain-related (e.g., level of criticality, compliance with standards) when introducing a specific approach in an industrial setting. For example, organizations that use model-based systems engineering (MBSE), e.g., to build automotive products, have a very different approach and mindset towards modeling than organizations where modeling is only done informally and not enforced by process descriptions or similar measures. Therefore, another objective of this work is to identify scenarios where modeling has proven successful, based on existing literature and our collective experience on the topic, and to distill future usage scenarios based on successful industrial practices.

Based on the work of \citet{bucchiarone2021future}, we regard the modeling success stories of MBSE, low-code, and informal modeling and add infrastructure-as-code as an additional success story. To gain a better understanding of informal modeling, we differentiate semi-formal modeling from it. Overall, we are thus regarding five success stories in this paper:
\begin{itemize}
    \item {\bf Infrastructure-as-Code (IaC)} aligns with the definition of modeling as the creation of system representations, where the models are embedded within the code itself. According to Madni et al.~\cite{Madni2018IaC}, a model can take various forms, including modeling languages, algorithms, equations, and parametric curves. This perspective offers a unified approach to both modeling and programming, suggesting that ``programs are models,'' where code is viewed as a less abstract, textual model \cite{Haxthausen2016IaC}. In this context, users are not necessarily aware that they are modeling. Domain-Specific Languages (DSLs) are mostly implemented as textual languages and take advantage of features of the usual tools/IDEs for programming, like automation and version control. Although these languages are not agnostic of the specificities of different tools, one might argue that IaC is already an abstraction with inherent structure and intentionality, reflecting a specific system deployment.
    We add IaC as it is a success story for the use of domain-specific languages, especially in a programming context.
    \item {\bf Low-code} is an emerging paradigm combining modeling and graphical programming. There are claims that Low-code is not MDE according to \cite{di2022low}, or that it is \cite{bucaioni2022modelling}. In this work, we follow the latter. Low-code platforms are modeling environments in which a user combines different pre-defined building blocks into an overall workflow. They are often used to model and automate simple processes or to create dashboards to visualise data. Examples of such platforms are Node-RED\footnote{\url{https://nodered.org}} or Outsystems\footnote{\url{https://www.outsystems.com/}}.
    \item {\bf MBSE in the Automotive Domain} uses models intensively for systematically developing and analyzing complex automotive systems architectures with dedicated modeling frameworks. In this domain, the most commonly used languages are UML and SysML to represent the structure of the software architecture at a high level of abstraction and to describe the system's behaviour via state machines. Stateflow/Simulink models are also used to represent subsystems which involve I/O and control design.
    \item {\bf Informal Modeling} produces models which are sketches with no or little structure that have no direct or automatic translation to code. The main objective is to gain a collective understanding of some problem domain, while engaging in intensive communication. 
    \item {\bf Semi-formal Modeling} produces models which are possibly created with CASE tools or diagrammatic tools such as draw.io, sometimes with limited automatic analysis or code generation capabilities.
\end{itemize}

The primary objective of this expert voice is to introduce factors that contribute to MX and their interrelations in the four identified modeling success stories. Moreover, it outlines how the identified factors impact the modeling success stories. This work results from the week-long GI-Dagstuhl Seminar on ``Human Factors in Model-driven Engineering'' attended by researchers and practitioners experts in the topics of model-driven engineering and human factors~\cite{dagstuhl23473}. The ultimate objective of this work is to contribute to designing and developing better MDE tools by understanding the MX factors that ultimately hinder MDE adoption.

This paper is structured as follows. In Section~\ref{sec:workflows}, we define and detail the five modelling success stories mentioned before. In Section~\ref{sec:MX_Factors}, we introduce the relevant factors that influence MX. In Section~\ref{sec:Factors_in_Workflows}, we analyze the five success stories in light of the factors identified. Finally, in Section~\ref{sec:conclusion}, we discuss the results and outline the research challenges and opportunities related to MX. 

\section{Selected Modeling Success Stories}
\label{sec:workflows}

\citet{bucchiarone2021future} introduce three modeling success stories as diverse but archetypal instances of successful applications of modeling. Due to their diversity, they exhibit a broad set of characteristics that influence the modeler's experience differently. Therefore, it is important to elicit the characteristics of each success story and analyze the extent to which they influence MX. As stated above, we also consider infrastructure-as-code as a modeling success story and consider it as such in the following.

Table~\ref{tab:modeling-workflows} presents the characteristics that define each modeling success story. These characteristics help to understand the differences and commonalities of the identified success stories. At this stage, we do not consider the list of characteristics to be complete but sufficient to show that the success stories are different enough to warrant a separate treatment.

\afterpage{%
    \clearpage%
    \thispagestyle{empty}%
    \begin{landscape}%
{
    \footnotesize
    \newcommand{\tabitem}{~~\llap{\textbullet}~~}
\begin{longtable} {@{}P{1.7cm}P{2.5cm}P{2.5cm}P{3cm}P{2.5cm}P{2.5cm}@{}}
\caption{Overview of the characteristics of the Modeling Success Stories based on \citet{bucchiarone2021future}}\label{tab:modeling-workflows}\\
\toprule
    & \textbf{IaC}
    & \textbf{Low-Code}
    & \textbf{MBSE}
    & \multicolumn{2}{c}{\textbf{Informal Modeling}}
    \\
    & \textbf{Helm Charts}
    &
    & \textbf{Architecture Modeling (AUTOSAR)}
    & \textbf{Informal}
    & \textbf{Semi-Formal}
    \\ \midrule
\endhead
 \textbf{Domain}                                
    & ``Online'' software services 
    & Any
    & Automotive
    & Any
    & Software
 \\ \midrule
\textbf{Goal(s)}                                     %
    & Automated deployment and testing of software services
    & Simplified app development
    &
    (Mandatory) Compliance (e.g., ISO 26262). Formalizing interfaces fensuring compatibility.
    Improving quality, reducing time-to-market, reducing cost.
    & Unobstructed creative flow of ideas
    & Documenting an architecture                 \\
\midrule
\textbf{Tooling}
    & Helm, Kubernetes, VS Code, git
    & Node-RED, Outsystems, Microsoft PowerBI
    & Heavyweight, extensive expert tooling ecosystem
    & Analog (paper, whiteboard) or Digital (touchscreen, \ldots)          
    & CASE tools (Flexisketch) or diagramming tools (PlantUML) \\
\midrule
\textbf{What is being modeled?} & Parameterized deployment of containers on Kubernetes cluster
                                & Data flows, business processes including UI and business logic
                                & Logical \& functional system architecture, using components, interfaces, connections. Deployment (HW nodes and bus systems), OS configuration, \ldots
                                & Abstract and concrete concepts. Early designs.
                                & Software (UML)                                           \\
\midrule \midrule
\textbf{Stakeholders}                               
    & Developers, Admins, DevOps engineers
    & Domain Experts, professional and citizen developers
    & Domain experts
    & Anyone
    & Software developers, architects                          \\
\midrule
\textbf{Human activities} & Specification, documentation, reviewing and manual verification
                          & Specification, documentation, reviewing and manual verification
                          & Specification, documentation, reviewing and manual verification
                          & Brainstorming, problem understanding, explaining
                          & Same as informal + documentation \\
\midrule
\textbf{Automated (machine) activities} 
    & Deployment \& testing of software services 
    & Data integration, code generation or runtime environment, setup and deployment
    & Simulation, testing, analysis (e.g., timing, compatibility), code generation 
    & N/A 
    & Code generation (limited) \\
\midrule \midrule

\textbf{File Format}         & Lightweight textual notation
                             & Textual (e.g., JSON) or binary
                             & XML
                             & Graphical
                             & XML or lightweight textual notation \\
\midrule
\textbf{Persistence}                    & Persistent, versioned
                                        & Persistent, versioned
                                        & Persistent, simple lock-based versioning (PLM)
                                        & (Analog) often discarded, erased, forgotten, \ldots
                                        & More often checked into version control \\
\midrule
\textbf{Re-use}                         & Yes (Helm: Packaging system)
                                        & Yes
                                        & Yes (PLM)
                                        & No
                                        & Yes (PlantUML: "include" command)
\\ \midrule
\textbf{(Perceived) advantages}         & Automation, reuse, git-integration
                                        & Speed, visual modelling, reuse, collaboration
                                        & Interfaces to Behavioral Specification (SimuLink). Time/Cost-Saving Generate OS. Efficient Code (e.g., communication via bus vs. communication via shared memory). Break vendor lock-in of HW(ECU) and SW.
                                        & High expressiveness, freedom and creativity. Low barrier of entry.
                                        & git-integration (PlantUML), clean, agreed-upon notations, multi-view. \\
\midrule
\textbf{Difficulties or limitations}                & Highly specialised tooling, quality assurance 
                                                    & Limited customization, security, vendor lock-in
                                                    & Complex tooling with poor scalability, collaboration and tool integration.
                                                     & Loss of context (speech, ...) over time. Poor persistence and traceability of analog sketches. Touchscreens often have poor responsiveness / usability.
                                                     & Less expressive (than informal), and limited analysis possibilities \\
\bottomrule
\end{longtable}
}
\end{landscape}
\clearpage%
}

The mapping in Table~\ref{tab:modeling-workflows} shows that the chosen success stories already provide a significant diversity in terms of the characteristics we have selected. This means they provide a sufficient difference regarding what modeler experience means in each of them and, ultimately, how they impact the factors of modeler experience as we describe in Section~\ref{sec:Factors_in_Workflows}.

\section{Factors of Modeler Experience}
\label{sec:MX_Factors}

In this section, we first describe the methodology we have followed to identify a set of factors that affect the modeler's experience. We then present these factors grouped by inherent factors, technical factors, and non-technical factors.

\paragraph*{Methodology}
We conducted a focus group with ten modeling experts from academia and industry to better understand the factors contributing to MX. The starting point of the focus group was a discussion of modeling success stories that can be observed in the industry. These success stories were defined in smaller groups consisting of two to three focus group participants. Each group identified several characteristics they deemed relevant for the success story described in Section~\ref{sec:workflows}.

While these characteristics were useful to distinguish the success stories and are used in Section~\ref{sec:workflows}, the group quickly realized that they are not suitable as descriptors of modeler experience since they are not focused on the modeler. Therefore, as a second step, the group engaged in a discussion that yielded factors more tailored towards the modeler: required training, maintainability, immediate benefits, integration in the programming ecosystem, and reduced friction between modeling and programming. Upon further reflection, the group identified that some of them were goals and others were very hard to measure, even qualitatively.

Therefore, the group engaged in a brainstorming session using the ``1-2-4-all'' technique: first, individuals brainstormed relevant factors on their own; second, two individuals came together as a group, discussed their respective factors, and consolidated them into a new list; third, four participants got together and consolidated again; and finally, the entire group discussed the different lists of factors and agreed on a final list of factors.

\begin{figure}
    \centering
    \includegraphics[width=0.9\textwidth]{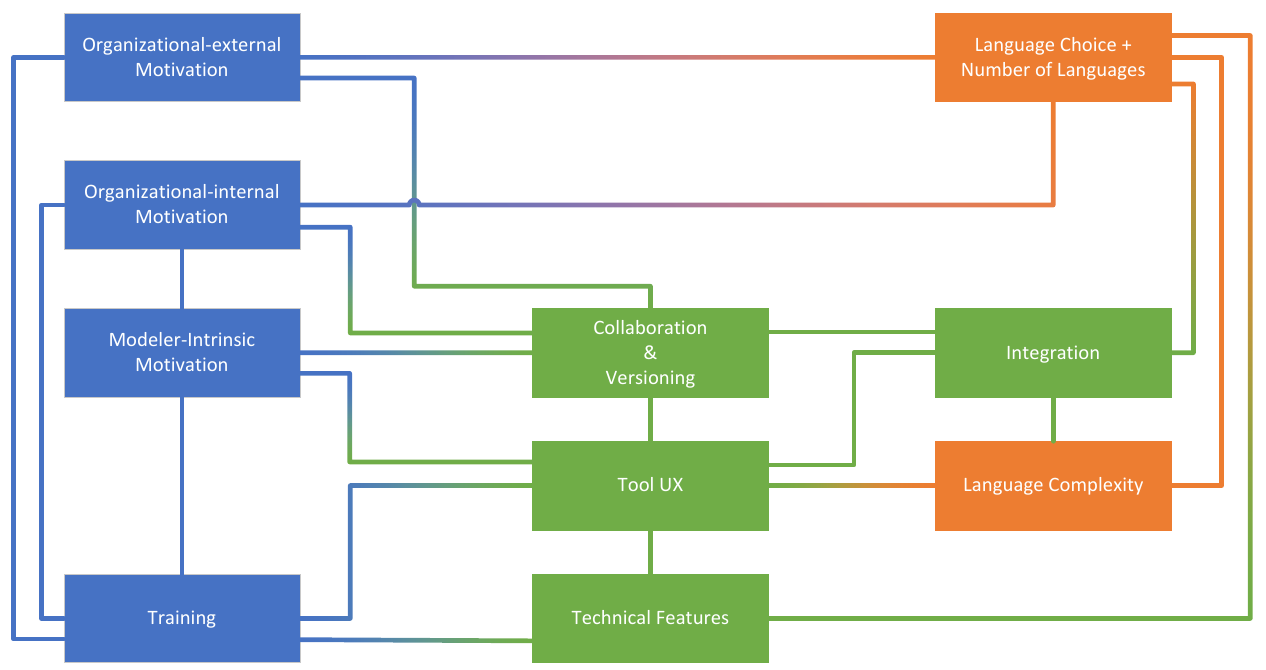}
    \caption{MX factors and their relations}
    \label{fig:MX_Factors}
\end{figure}

As an additional step, the focus group participants then discussed how the different factors relate to each other (cf.~Figure \ref{fig:MX_Factors}). While originally trying to identify positive and negative influences, this idea was abandoned at one point in favor of a more generic notion of relationship. It is unnecessary to distinguish cases in which a factor can positively or negatively influence another.

The result of these intense discussions during the focus groups is the set of factors described below.

\paragraph*{Inherent Factors}
An inherent factor is based on the characteristics of the problem to address. The complexity of the language is inherent in the complexity of the problem. Language choice is also based on the domain and, ultimately, on the problem that needs to be solved. The chosen language needs to be able to address that problem and needs to have the necessary complexity to describe relevant issues in the problem domain. 
\begin{itemize}
    \item \textbf{Language Complexity} is a measure of effort to solve a modeling problem using a modeling language.
    Perceived language complexity has been listed as an impacting factor in~\cite{abrahao2017user}.
    In their seminal work, France et al.~\cite{France2007Seminal} highlight the challenge of \emph{Managing Language Complexity} faced by practitioners. Researchers have subsequently employed this classification to assess and prioritize challenges practitioners encounter in software modeling~\cite{Ozkaya2020Survey}.
    
    \item \textbf{Language Choice} is the process of selecting, extending, or defining a set of modeling languages that provide suitable domain-specific abstractions. Vendor lock-in creates barriers to MX since it limits interoperability. It should, therefore, be included as a criterion in the selection process. 
    Language choice is also influenced by the \textbf{system domain}, ``a sphere of knowledge, influence, or activity. The subject area to which the user applies a program is the domain of the software.''~\cite{evans2014domain}
    According to a literature review~\cite{kalantari2022slr}, the language used directly influences MX. Modeling Languages usually also cover multiple \textbf{view points}~\cite{gonzalez2013defining}.
    The need for modeling viewpoints was emphasized for practitioners and is evaluated based on its support for multiple viewpoints and large viewpoint management~\cite{Ozkaya2019Ltr}.
\end{itemize}

\paragraph*{Technical Factors}
These factors concern the way modelers work with the chosen languages and integrate them into their workflows. Specifically, we identified four technical factors:
\begin{itemize}
    \item \textbf{Integration} is the conceptual and technical capability of the modeling approach to fit into existing development processes and development platforms.
    Whittle et al.~\cite{Whittle2015Taxonomy} categorize this issue under ``Practical Applicability/Challenges of Applying Tools in Practice.'' This is captured in two sub-factors: 1) chaining tools together, addressing the ease or difficulty of using multiple tools for end-to-end functionalities, and 2) flexibility of tools, evaluating a tool's adaptability to various processes, tools, and working methods without imposing strict processes or requiring additional tools~\cite{Whittle2015Taxonomy}.
    \citet{Mohagheghi2013study} report that three of four companies mentioned significant efforts to integrate MDE into the existing processes. The domain was identified as a contributing factor towards adoption. Participants voiced that MDE seems most beneficial for bigger companies, projects, or companies with product lines or similar projects. The vision paper \cite{combemale2020Devops} identifies integration into DevOps workflows as an important step towards adopting MDE. They focused specifically on the domain of cyber-physical systems, which would include the automotive industry. 
    
    \item \textbf{Tool UX}: ``A person's perceptions and responses that result from the use and/or anticipated use'' of the modeling tool~\cite{iso9241-210}.
    \textit{Ease of use} and \textit{maturity of tools} are two aspects mentioned by \citet{Mohagheghi2013study}. Both aspects can be categorized as Tool UX, where tool maturity is likewise closely related to integration and technical features. Interestingly, the authors found that companies with lower adoption of MDE considered maturity of tools to be worse than those with higher adoption. %
    Some participants suggested enhancing the UI of modeling software to improve Tool UX.
    In their systematic literature review, \citet{kalantari2022slr} classified and documented MX issues into five distinct categories: utility, usability, reliability, emotional, and marketing. Utility and usability are related to ease of use, while reliability corresponds with the maturity of tools.
    
    \item \textbf{Versioning} is the ability to track and merge model changes, facilitating model maintenance, and is one of the main enablers of \textbf{Collaboration}, as stated by \citet{pietron2020collaborative}. This suggests that versioning is also an important factor of MX.
    The inadequacy of version management support has been identified as a drawback of Model-Based Engineering (MBE) tools within the embedded systems domain~\cite{Grischa2014Embedded}. The technical challenge of combining versioning with blended modeling is addressed by \citet{exelmans2022optimistic}.
    \textbf{Collaboration} typically alternates between two modes: asynchronous collaboration, where the work is divided, and synchronous collaboration, to obtain a shared understanding~\cite{conway2008holistic}.
    In the literature, support for collaboration has consistently emerged as a desired attribute of modeling tools for practitioners~\cite{Ozkaya2019Ltr,Badreddin2018trends, BordeleauLRST17,david2023collaborative,franzago2017collaborative}. In the more general field of software engineering, collaborative tools positively impact productivity~\cite{hidayanto2014impact,ur2020use}.

    \item \textbf{Technical Capabilities} result from the combination of modeling language and modeling tool. They describe how the models can be used in the development process, e.g., reason about properties of the modeled system via simulation or formal analysis or to generate downstream artifacts. For example, Liebel et al.~\cite{DBLP:journals/sosym/LiebelMTLH18} present results from a survey about the technical capabilities used in model-based engineering in the embedded systems domain. Störrle~\cite{DBLP:conf/ease/Storrle17} presents results on a survey for which purposes models are used when modeling informally.

\end{itemize}

\paragraph*{Non-technical Factors}

These factors relate to modeling outside the problem domain and its technical use. Specifically, we identified four non-technical factors:

\begin{itemize}
    \item \textbf{Modeler-Intrinsic Motivation} are those factors that lead modelers to use modeling, in particular, the perceived benefits and positive emotions they experience. \textit{Intrinsic motivation} is the inherent motivational source that drives individuals to engage in activities they personally find compelling. This stands in contrast to extrinsic motivation, where the incentive originates from an external source rather than the inherent appeal of the task~\cite{Ryan2000Intrinsic}. In software development, empirical research suggests that intrinsic motivation significantly predicts developers' overall experience~\cite{Kuusinen2016Intrinsic}. On the other hand, a lack of perceived benefit of modeling has been noted as one of the key mindset barriers for adopting MDE~\cite{kalantari2023unveiling}.
    
    \item \textbf{Organization-Intrinsic Motivation} are the factors inside the organization that lead the organization and the modelers to adopt and foster modeling, in particular, because of perceived benefits for productivity, product quality, cost, or collective well-being. The benefits of software modeling practices have been widely discussed in the literature. A survey on modeling practices in the embedded software industry underscores cost savings, shortened development time, reusability, and improved quality as key motivations for adopting MDE~\cite{Akdur2018Survey}. An empirical assessment noted the benefits of MDE in communication and control within a project~\cite{Hutchinson2011Empirical}. The assessment also noted that MDE adoption is affected by culture, expertise, and evangelism within the organization. MDE adoption requires organizational changes, notably, the need for a modeling `champion' and carefully choosing initial projects for applying MDE~\cite{Hutchinson2011Empirical}. \citet{Vogelsang2018Embedded} noted the importance of managing expectations when adopting MDE.
    
    \item \textbf{Organization-Extrinsic Motivation} are the factors outside the organization which influence adoption, maturity, and approach of/to modeling, e.g., existing standards, regulations, tool availability and maturity, or customer demands. Adhering to regulations is one of the strong drivers for MDE adoption in the embedded systems industry~\cite{Vogelsang2018Embedded}. The Unified Modeling Language (UML) has been cited as \emph{de facto} standard for software modeling, with a lot of tools available that support not only model creation and code generation but also viewpoint management, verification, etc.~\cite{Ozkaya2019Ltr}. However, it has been noted that existing research on modeling does not address quality issues reported in industrial context~\cite{giraldo2018considerations}.

    \item \textbf{Training} includes all factors that are related to the skills and knowledge of modelers in using the selected modeling languages and the tools used to create, manipulate, analyze, and use the models. Training and education have been mentioned as an important factor affecting MDE adoption~\cite{Hutchinson2011Empirical}.  Insufficient training resources and support, on the one hand, \cite{kalantari2023unveiling}, and the substantial effort required for developer training on the other \cite{Liebel2018Embedded, Vogelsang2018Embedded}, are two significant factors in this context. \textit{Lack of fundamentals in MDE} and \textit{education issues} have been noted as major current problems in MDE~\cite{mussbacher2014relevance}. 
    Moreover, perceived competence, defined as an individual's subjective judgment regarding their own skills and performance, is identified as a crucial factor contributing to intrinsic motivation \cite{Kuusinen2016Intrinsic}. This recognition underlines the interconnected nature of factors in this domain, emphasizing that effective training impacts proficiency and influences individuals' intrinsic motivation in model-driven practices.
\end{itemize}

An important aspect of the different motivations is the willingness to take risks on the individual and organizational level. A small organization might be more willing to take a risk with its modeling approach and the tools used than a larger organization because it is more driven by the need to innovate and cannot afford a less risky but more expensive solution.

\section{Applying MX to the Modeling Success Stories}
\label{sec:Factors_in_Workflows}

In this section we highlight how the different MX factors apply to the modeling success stories introduced previously. We structure each section according to the different groups of factors. Table~\ref{tab:mapping-factors-workflows} also shows the importance of the different factors to the different sucess stories.

\afterpage{%
    \clearpage%
    \thispagestyle{empty}%
    \begin{landscape}%
\begin{center}
\begin{table}[tb]
\caption{Mapping of the identified factors to the modelling success stories. The table also shows the importance of the different technical and non-technical factors for the individual success story.}
\label{tab:mapping-factors-workflows}
\begin{tabular}{@{}P{3.2cm}P{2.2cm}P{1.7cm}P{2.1cm}P{2.2cm}P{2.2cm}@{}}
\toprule
                & \textbf{Infrastructure as Code} & \textbf{Low-Code} & \textbf{MBSE}  & \textbf{Informal Modelling} & \textbf{Semi-formal Modelling} \\
\midrule
\multicolumn{6}{@{}l}{\textbf{Workflow-Inherent Factors}}\\
\midrule
Language Complexity               & Medium                & Medium              & High                & Low         & Low/Medium    \\
Language Choice                   & Medium                & Restricted          & Restricted          & Free        & Free       \\
\midrule
\multicolumn{6}{@{}l}{\textbf{Technical Factors}}\\
\midrule
Integration                       & $++$                 & $++$                 & $-$         & $+$        & $++$ \\
Tool UX                           & $+$                  & $++$                 & $++$        & $+$        & $-$ \\
Versioning \& Collaboration       & $++$                 & $++$                 & $+$         & $+$        & $++$ \\
Technical Features                & $--$                 & $+$                  & $-$         & $+$        & $++$ \\
\midrule
\multicolumn{6}{@{}l}{\textbf{Non-technical Factors}}    \\
\midrule
Modeler-Intrinsic Motivation      & $++$                 & $-$                  & $++$        & $++$       & $-$ \\
Organisation-Intrinsic Motivation & $+$                  & $+$                  & $-$         & $-$        & $+$ \\ 
Organisation-Extrinsic Motivation & $-$                  & $--$                  & $--$        & $--$       & $+$\\
Training                          & $++$                 & $+$                 & $-$         & $+$        & $++$\\
\bottomrule
\end{tabular}
\end{table}
\end{center}
\end{landscape}
\clearpage%
}

\subsection{Infrastructure as Code}
\noindent \textbf{Inherent Factors} Infrastructure as Code is typically used at the beginning of the development process when an infrastructure is needed to execute a developed software. This infrastructure is often managed by a single or only a few DevOps Engineers and evolves as the development process continues. 
The language depends on the chosen cloud provider or container orchestration tool, so the choice is limited. While it is simple to configure a CRUD web application, due to re-usability, community support, and extensive documentation, configuring specialized distributed systems may require more complex features of IaC languages.

\noindent \textbf{Technical Factors} As IaC is usually written in an IDE (with additional language support), it directly benefits from many features such as versioning, auto-completion, formatting, and testability. IaC is, therefore, highly integrated into the development workflow. Command line interfaces or simple graphical interfaces are then used for provisioning and setting up infrastructure from IaC-configurations, which provides developers with a familiar tool's UX. 

\noindent \textbf{Non-Technical Factors} Infrastructure as code is adopted by DevOps engineers to reduce repeated work, improve debugging capabilities, and improve consistency across similar software systems. Organizations further aim to reduce costs, gain flexibility, and improve documentation of their systems. However, depending on previous workflows and infrastructure management, the change to IaC can be demanding and requires training for operations engineers and developers.

\subsection{Low-Code}
\noindent \textbf{Inherent Factors} Mature low-code platforms support the entire application lifecycle management. They do use proprietary modeling languages, though, so language choice is limited by the chosen tool. Different platforms specialize in different use cases: Node-RED, e.g., is marketed as a tool for data processing and visualization for embedded systems whereas Microsoft's PowerBI platform is geared towards business process automation and visualization. With these use cases come different feature sets, different language complexity, different levels of extensibility and different support options that modelers have to take into consideration.

\noindent \textbf{Technical Factors}  Low-code platforms provide their own development environment. In many cases, these environments include collaboration and versioning features as well as features to automatically run, test, provision, and deploy the application using cloud-native concepts. They are separate from other IDEs developers might be using for other parts of the system, however. Usability of these tools is generally quite good~\cite{pinho2023usability} since the visual aspect of the environment is one of the most important features and many low-code platforms are specifically aiming at non-professionals. 

\noindent \textbf{Non-Technical Factors} The switch to low-code platforms is often motivated by organization-wide decisions, e.g., to move to Microsoft PowerBI to modernize legacy systems. We cannot discern situations in which organisation would be pushed to adopt such systems due to external circumstances. Depending on the use case, the barrier of entry for such platforms can be significant, but documentation and training material are exhaustive and there are a variety of training options.

\subsection{MBSE}

\noindent \textbf{Inherent Factors} The modeling success story in the Model-Based Systems Engineering domain is characterized by extensively using highly sophisticated and standardized modeling languages. Examples are AUTOSAR for modeling the hardware/software architecture, and Matlab/Simulink/Stateflow for modeling the software's behavior. Due to the ability to generate production code from the models, the modeling languages are highly complex and provide many technical features. The language choice in that domain is restricted as complex value networks between OEMs and suppliers require extensive artifact exchange and, thus, language standardization.

\noindent \textbf{Technical Factors}  In addition to the code generation capabilities, the tools also provide extensive analyses and simulation capabilities. Furthermore, the tools support versioning and integrating software artifacts, e.g., an OEM integrates components from various suppliers. Tool UX is often rather low as the tools are expert and niche tools in addition to the complex languages and the number of technical features. 

\noindent \textbf{Non-Technical Factors}  
The restrictions of the domain highly influence the non-technical factors of the modeling experience in the MBSE success story. Modeling languages and the modeling success story are restricted by organization-extrinsic motivation, e.g., standards or clients require companies to use certain modeling languages and corresponding tools. Furthermore, the company-internal and company-external collaboration is dependent on using the same modeling languages and conforming modeling tools.

\subsection{Informal and Semi-Formal Modeling}

\noindent \textbf{Inherent Factors} Informal Modeling mostly occurs early in the design process, where conformance to any standardized notation is of minimal concern, as opposed to (creative) expressiveness and a free flow of ideas. Often, the created sketches evolve into (semi-)formal models later on.

\noindent \textbf{Technical Factors} Much of the informal modeling occurs with analog media such as pen and paper or whiteboards. Although digital alternatives (e.g., electronic whiteboards) exist, their adoption rate remains low in practice, despite technical advantages such as automated persistence and remote collaboration. We think this remains due to poor Tool UX (e.g., responsiveness, viewing angles, accuracy, \ldots) of these solutions compared to analog media, and we conclude this factor is of utmost importance for informal modeling.

\noindent \textbf{Non-Technical Factors} Informal modeling, and its low-tech nature, comes natural to engineers, and has been done even since before it was considered a form of `modeling'. Workforce is intrinsically motivated, and this motivation does not depend on organization-intrinsic or -extrinsic factors. It is often a collaborative activity.

\subsection{Comparing the success stories}

When comparing the different modeling success stories, we see they are in different places on the spectrum of MX factors. We illustrate the diversity of modeling success stories w.r.t.~language complexity and technical capability of the tooling as an example. We believe there are different sweet spots for different modeling workflows: for instance, model-based systems engineering reaps benefits from being formal and, therefore, having high language complexity; this requires tools with high technical capability, which are also very complex to use. Informal modeling, on the other hand, uses languages with very low complexity and requires tools with less technical capabilities, e.g., to reason about models. Both modeling success stories are in stark contrast to UML and its tools in the early days when language complexity was high, but at the same time, rather informal, and the tools had low technical capability.  

When using other factors, the different success stories will be positioned differently. For instance, when contrasting language choice and organisation-external motivation, MBSE with a high motivation and few available languages poses fewer issues than informal modelling where languages are often made-up on the spot and syntax might not be understood between different companies nor even across teams. This illustrates that our factors are able to capture interesting trade-offs and that a more systematic investigation of these trade-offs in future research is necessary.

\section{Conclusion and Future Work}
\label{sec:conclusion}

We defined and explored the concept of Modeler Experience (MX) in the context of different  modeling success stories: infrastructure-as-code, model-based systems engineering (MBSE), and informal and semi-formal modeling. MX, which encompasses factors such as usability, motivation, integration, and language complexity, highlights the dynamics between practitioners and modeling tools. One of the contributions of this paper is the delineation of technical and non-technical MX factors and the characterization of these factors in different modeling success stories. In addition, we show how the proposed MX factors differ between those modeling stories.
By examining these sucess stories in the light of MX factors, the paper underscores the contextual nature of modeling practices and the need for tailored approaches to effectively address modelers' needs.

Although this paper lays a solid foundation for understanding MX, there are still many potential research paths to be explored in the future:

The interplay between MX factors and the success of modeling activities suggests the need for a holistic approach to tool design and workflow integration. Moving forward, research should focus on empirically validating the identified MX factors across diverse contexts and exploring the impact of different workflows on adopting and sustaining modeling practices.

Further empirical studies are essential to quantify the relative importance of each MX factor and how they interact with one another. These studies could take the form of longitudinal research within organizations that are transitioning to or already utilizing model-driven approaches, as well as experimental studies comparing different tool sets and methodologies. Additionally, there is an opportunity to conduct cross-sectional analyses comparing MX across domains, such as automotive, aerospace, healthcare, and finance, where modeling plays a crucial role.

Another promising direction is the development and refinement of modeling tools that incorporate the principles of human-centered design. By focusing on the end-user experience, tool developers can create more intuitive interfaces, improve the integration with existing development ecosystems, and enhance collaborative features. This direction also calls for an iterative design process, where feedback from modelers is continuously incorporated into tool enhancements.

Case studies can provide in-depth insights into the practical challenges of addressing MX in specific contexts and the benefits of modeling in industry. By examining specific cases in detail, researchers can identify best practices, common pitfalls, and innovative uses of modeling tools that may not be evident through other research methods.

In conclusion, by continuing to explore and address the various dimensions of MX, we will serve both academic research and practical tool development, guiding efforts toward powerful models and tools that are a pleasure to use, thereby enhancing productivity and satisfaction for modelers worldwide.

\backmatter

\bmhead{Acknowledgments}

This work is an outcome of the GI-Dagstuhl-Seminar 23473 ``Human Factors in Model-driven Engineering''. The authors are grateful to the Leibniz Center for Informatics and the seminar's organizers for the opportunity to collaborate and produce these results. Abrahão’s contributions were partially supported by the State Research Agency (AEI) under the UCI-Adapt project (PID2022-140106NB-I00). Amaral' contribution was partially supported by FCT - Fundação para a Ciência e Tecnologia (Grant Number: 2021.08371.BD).

\bibliography{sn-bibliography}%

\end{document}